\documentclass[journal=jpcb,manuscript=article]{achemso}

\usepackage{chemformula} 
\usepackage[T1]{fontenc} 
\usepackage{float}
\usepackage{subfig}

\usepackage{setspace}
\usepackage{times}
\usepackage{appendix}
\usepackage{bm}
\usepackage{graphicx}
\usepackage[font=small]{caption}
\usepackage{subcaption}
\usepackage[dvipsnames]{xcolor}
\usepackage{natbib}
\usepackage{array}
\usepackage{epsfig}
\usepackage{latexsym}
\usepackage{epstopdf}
\usepackage{amsfonts}
\usepackage{amsmath}
\usepackage{mathtools}
\usepackage{grffile}
\usepackage{multirow}
\usepackage[english]{babel}
\usepackage[autostyle]{csquotes}
\usepackage[export]{adjustbox}
\usepackage[normalem]{ulem}
\usepackage[colorlinks=true,linkcolor=blue,urlcolor=blue,citecolor=blue]{hyperref}
\usepackage{hyphenat}
\newcommand{\pKa}{\mathrm{p}K_{\mathrm{a}}}

\author{Mamta Yadav}
\affiliation{Computational Chemistry, Lund University, P.O.Box 124, S-221 00 Lund, Sweden}
\author{Clifford E. Woodward}
\affiliation{School of Physical, Environmental and Mathematical Sciences University College, University of New South Wales, ADFA Canberra ACT 2600, Australia}
\author{Jan Forsman}
\email{jan.forsman@compchem.lu.se}
\affiliation{Computational Chemistry, Lund University, P.O.Box 124, S-221 00 Lund, Sweden}

\title{A Boosted Energy Extraction from the CapMix Process by Grafting with Titratable Polymers}

\keywords{CapMix, Blue Energy, simulations, cDFT, electrolytes, titration}

\begin{document}



\begin{abstract}
Salinity gradient energy offers a sustainable route to convert ionic
chemical potential differences into usable power.  Capacitive mixing
enables this conversion without membranes, but suffers from limited
ion regulation at electrode interfaces.  Here we show that by grafting
electrode surfaces with titrating polymers, the performance can be
substantially improved.  Using Grand Canonical Monte Carlo simulations
with exact image-charge Ewald summations, we demonstrate how the
coupled effects of ion adsorption and charge regulation in response to
an external potential can be harnessed.  Grafted electrodes are shown
to deliver substantially more
energy relative to bare surfaces, driven by charge regulation effects
that exploit the pH difference that typically exists between rivers and
the ocean.  While the effect is in principle maximized at high grafting
densities and moderate chain lengths, the performance is fairly robust
to variations of these parameters, within reasonable bounds.
Complementary classical polymer Density Functional Theory calculations
confirm these trends, validating the
mechanistic framework.  This work also establishes a practical
approach to harvest electrical energy during wastewater neutralization,
where acidic (or alkaline) effluents serve as complementary
reservoirs, and offers a promising strategy to couple environmental
remediation with renewable energy recovery.
\end{abstract}


\section{Introduction}\label{IN}
\setcounter{equation}{0}
\renewcommand{\theequation}{1.\arabic{equation}}

Human activities have caused significant environmental stress, making
the transition toward renewable energy sources increasingly urgent
\cite{Chu2016}.  Meeting the growing global energy demand requires the
development of diverse and sustainable technologies \cite{Chu2012}.
While wind \cite{JOSE2007,TAN2022} and solar power \cite{KANNAN2016}
currently dominate the renewable energy sector, salinity gradient
energy---commonly known as blue energy---has attracted increasing
attention in recent years.  This source of energy exploits the mixing
of freshwater with seawater at coastal interfaces, where a 
natural salinity gradient arises
\cite{Yip2016,ALVAREZSILVA2016}.  Brines from natural salt lakes,
coal-mining activities, or concentrated salterns offer alternative
sources for salinity gradient energy when mixed with seawater, thereby
reducing the reliance on freshwater resources
\cite{ZHU2024,TUREK2008,tufarscadv2014}.  The free energy released
from this gradient can, in principle, be harvested and converted into
usable power.  When freshwater mixes with seawater at estuaries, about
$2\,\mathrm{kJ}$ of free energy per litre of river water is
dissipated.  On a global scale, this corresponds to a potential of
approximately ${\sim}\,2\,\mathrm{TW}$, making it one of the largest
marine-based renewable energy resources \cite{WICK1978,Kuleszo2010}.
If not harvested, this significant amount of energy is simply lost
during the natural mixing process.

Efforts to extract blue energy have mainly focused on membrane-based
technologies such as Pressure Retarded Osmosis (PRO)
\cite{Yip2012,HAN2013} and Reverse Electrodialysis (RED)
\cite{Post2007,Yip2014}.  While these convert salinity gradients into
electricity using semipermeable or ion-selective membranes, their
performance is limited by high cost, energy-intensive pumping, and
fouling \cite{straub2016,VERMAAS2013}.  These drawbacks have driven
interest in Capacitive Mixing (CapMix)
\cite{BROGIOLI2013,debrogioliEE2012}, a membraneless alternative that
harvests the free energy of mixing through cyclic ion adsorption and
desorption on porous electrodes.  Brogioli \textit{et al.}\
\cite{debrogliePRL} first demonstrated this concept, and La Mantia
\textit{et al.}\ \cite{LaMantia2011} later enhanced it via charge-discharge 
cycles between seawater and river water.  Ongoing research
now focuses on improving efficiency through advanced electrodes
\cite{MARINO2014}, optimised architectures \cite{Sales2010}, and
hybrid designs such as the Capacitive Donnan Potential (CDP) system,
which outperforms the classical Capacitive Double Layer Expansion
(CDLE) approach but still falls short of industrial benchmarks
\cite{ZOU2022}.

The energy harvested in CapMix systems depends on the relationship
between surface potential and charge, which is governed by the ionic
distribution within the Electric Double Layer (EDL).  Recent studies
indicate that grafting electrodes with polyelectrolytes can markedly
modify the ion distribution and the electrode response
\cite{HU2024,Chen2025}.  Polyelectrolytes, consisting of ionizable
polymer chains, exhibit charge states that strongly depend on the
surrounding pH, with weak polyelectrolytes showing pronounced
pH-sensitive dissociation \cite{geoghgen2022,Wang2008}.  Their
restricted conformations lead to preferential counterion adsorption,
creating inhomogeneous ion distributions near the surface.  The
tunable ionization of such grafted layers can thus be exploited to
harvest energy from pH gradients, for example during the neutralization
of industrial wastewater \cite{Culcasi2021}.  Since many waste streams
are either acidic or alkaline, coupling their treatment with energy
recovery offers a promising route toward simultaneous neutralization
and sustainable power \cite{Zaffora2020}.

Motivated by a study demonstrating that polyelectrolyte hydrogel-coated
membranes can achieve osmotic energy harvesting at
$5.06\,\mathrm{W\,m^{-2}}$ \cite{Zhang2020}, recent theoretical work
has investigated polyelectrolyte-coated porous electrodes.  These
studies primarily considered fixed-charge polyelectrolytes, showing
that grafting can enhance blue energy extraction by up to $45\%$
compared to bare electrodes \cite{Tian2024}.  However, the behaviour
of pH-responsive (titratable) polyelectrolytes, which allow tunable
charge through ionization, remains largely unexplored.
Moreover, most existing theoretical approaches rely on
classical density functional theory (cDFT) or related
approximations \cite{Tian2024,Jing2024}, typically neglecting
contributions from image charges, and with an approximate
treatment (if at all) of ion correlations.  Therefore, this work focuses on
investigating the effects of coating conducting electrodes with
titrating polyelectrolytes and their influence on salinity-gradient
energy harvesting.  Despite extensive theoretical efforts, simulations
of ionic systems near conducting electrodes remain limited due to the
complexity of image-charge treatments and slow
convergence \cite{Kondrat2023,Ballenegger2009}.  A grand
canonical scheme based on exact image-charge Ewald summations offers
an efficient alternative \cite{Janne2020}, which we here extend to
titrating polyelectrolytes and apply to the promising context of blue
energy extraction.  Furthermore, we have incorporated image-charge
effects, albeit within certain approximations, into the cDFT, and
systematically compared the theoretical predictions with our Monte
Carlo simulation results, demonstrating the consistency between the
two approaches.

\section{Model and Theory}

The extraction of salinity gradient energy through CapMix depends on
the cyclic charging and discharging of a porous electrode alternately
exposed to two electrolytes of different ionic strength and pH.  The fundamental
working principle can be found in several references
\cite{debroglieEE,BROGIOLI2013}.  Here, we consider a $4$-step
electrochemical cycle (analogous to the Carnot process for heat engines), where the role
of temperature difference is replaced by the
salinity change between seawater and freshwater
\cite{debrogliePRL,Boon2011}.

We denote the general seawater/freshwater system as the $aqueous$ phase,
which includes the electrochemical engine situated at the interface where salinity mixing occurs.
The CapMix cycle directs some of that mixing to occur through the engine, from which work is extracted.  
During the cycle heat is exchanged with the surrounding atmosphere in order to maintain the temperature $T$ 
of the aqueous phase. 

In the $1st$ step, an electrode is immersed in seawater
and has specifically chosen surface charge density  $\sigma = \sigma_i$, with a surface
potential of $\psi = \psi_{init}$.   The surface charge density is then increased to a higher value
$\sigma_f$ under an applied external potential, $\psi_{\mathrm{ext}}$.  
This can be approximately accomplished pragmatically by connecting the electrode to an external 
battery that maintains a potential equal to $\psi_{\mathrm{ext}}$. 
This step corresponds to ion adsorption at the electrode surfaces 
from the surrounding seawater and is assumed to take place in a {\em reversible} manner
so that the (absolute) surface potential increases quasi-statically from $\psi = \psi_{init}$ 
(at  $\sigma = \sigma_i$) to  $\psi_{\mathrm{ext}}$ (at  $\sigma = \sigma_f$). 
The $2nd$ step is the flushing of the seawater electrolyte by freshwater while maintaining the
surface charge.  The sudden decrease in ionic concentration reduces the
screening of surface charges, leading to a non-quasistatic jump in the surface
electrode potential from $\psi_{\mathrm{ext}}$.  
It is envisaged that the external battery is disconnected during this process.  
In the $3rd$ step, the electrode is assumed to 
discharge reversibly in freshwater until the surface potential is back at the 
value $\psi_{\mathrm{ext}}$, resulting in the transfer of ions from the electrode into the bulk freshwater environment.  
In a practical sence, this would involve reconnecting the battery.
The cycle is contrived so that the surface charge density at this point in the cycle is 
equal to its initial value, $\sigma_i$.  Finally, the $4th$ step, is the reverse of the $2nd$ step, and freshwater 
is flushed out by seawater at the fixed surface charge so that the system returns to the 
starting point of the cycle, with the surface potential equal to $\psi_{init}$. 
  
During this cycle, a number of cations, $\delta N_+$, and anions, $\delta N_-$, are transferred
from the seawater to the freshwater.  This increases the total entropy of 
the aqueous phase by an amount, $\delta S_{aq}$,
due essentially to ionic mixing.  The total heat $\delta Q$, absorbed by the 
aqueous phase through the interface with the atmosphere,
satisfies the Clausius inequality,   $\delta Q/T  \le \delta S_{aq}$, where $T$ is the 
ambient temperature.  The equality holds for equilibrium cycles, while for non-equilibrium 
cycles the total entropy increases.  A classic example is the
uncontrolled (spontaneous) mixing of two differently labelled ideal gases, which 
would not exchange any heat with a connected reservoir.
There is also an energy change at the completion of the cycle, denoted as $\delta U_{aq}$.  
We expect that $\delta U$ is positive, due to reduced electrostatic attractions in the more dilute freshwater component.
We have  $\delta U_{aq} = \delta Q - W$,  where $W$  is the total electrical work extracted by the electrodes. 
Thus we have in general that, $W \le T\delta S_{aq} - \delta U_{aq}$, where equality occurs for reversible cycles.  In the case
described above, the $2nd$  and $4th$ flushing stages are not reversible, so the work
performed is not maximal.   Given that the aqueous phase contains essentially two bulk reservoirs, 
the maximal amount of work, $W_{max}$, theoretically possible
is given by the negative free energy change upon dispersing the transferred ions from the bulk seawater
to bulk freshwater.  Thus,  $W_{max} = \delta N_+\Delta \mu_+ + \delta N_-\Delta \mu_-$.  Where, $\Delta \mu_\alpha$, is the
chemical potential difference for ions of type $\alpha$ in the seawater relative to the freshwater.

The net electrical work, $E$ , extracted (per unit electrode area) can be found by considering the graph of the electrode 
potential ($\psi$) vs. the surface charge density ($\sigma$), and then calculating the 
area enclosed by the cycle described above.  In steps $2$ and $4$, the potential $\psi$ jumps
vertically at constant $\sigma$, thus
\begin{equation}\label{eq1.1}
  E = \int_{\sigma_i}^{\sigma_f} \Delta\psi(\sigma)\,\mathrm{d}\sigma,
\end{equation}
where $\Delta\psi(\sigma)$ is the difference in electrode potential
between the low- and high-salinity cases at surface charge
density $\sigma$.  This cyclic variation of surface charge and
potential allows the conversion of the chemical potential difference
between seawater and freshwater into usable electrical energy.  

Our model electrode is assumed to be porous and perfectly conducting
(polarizable), with a surface charge regulated by an applied Donnan potential
(so-called ``$\Psi$-control'').  To reduce computational cost while capturing the
essential physics, we model the single porous electrode as a
planar slit.  This simplification allows us to relatively easily calculate the local ion distribution,
surface charge, and system energy, and the results
can be scaled by the total pore surface area to estimate the macroscopic
charge and the work generated.  The walls of the planar pore are separated by a width
$h$, which serves as a characteristic pore size.  
In this study, we assume that the electrode carries a positive
charge surface and is uniformly coated  with weakly acidic polymers \cite{Borukhov2000}.   

Application of a positive electric potential will promote proton dissociation 
in the grafted chains.  This results in negatively charged chains that act to screen the
positive surface charge.  This cooperative coupling
enhances the capacitance relative to the bare
electrode.  In contrast, if the same weakly acidic polymers were
grafted onto the cathode (negative potential), ionization would be
suppressed, rendering the polymer chains nearly neutral, with a
negligible impact on the capacitance.  For a complete cell, we
would recommend polymer grafting only at
the positive electrode, which still will provide a
significant overall improvement, as we shall demonstrate. 

The ionic solution is modelled using the restricted primitive model, where ions
are treated as charged hard spheres of diameter $d = 3\,\text{\AA}$ in
a uniform dielectric medium.  The ions are monovalent
(valency $Z_i = \pm 1$), while the solvent has a relative permittivity
$\epsilon_r = 78.5$ at $T = 298\,\mathrm{K}$.  Our treatment of just a single electrode
assumes that the cathode and anode separation is
large enough to neglect mutual interfacial effects.  
The titrating polymers are modelled
as freely jointed chains of connected hard-spheres having the
same diameter as the ions and a fixed bond length of $4\,\text{\AA}$.
All chains have the same degree of 
polymerization, $R$.  Unless otherwise stated, we have set  $R = 6$.
One end monomer of each chain is grafted to the electrode surface
giving a uniform surface charge density of grafted chains, denoted by $\rho_g$. 
Unless otherwise specified, we have set $\rho_g = 0.032\,\text{\AA}^{-2}$.
 
The pair interactions between the charged
hard spheres in our model are defined as:
\begin{equation}
  \beta u(r) =
  \begin{cases}
    \infty,                       & r < d,      \\[2mm]
    \ell_B \dfrac{Z_i Z_j}{r},   & r \ge d,
  \end{cases}
\end{equation}
where $r$ is the centre to centre distance and the length scale of 
the electrostatic interaction is set by the
Bjerrum length $\ell_B = e^2/(4\pi\epsilon k_B T)$.  Here, $e$ is
the elementary charge, $k_B T$ is the thermal energy, and
$\epsilon = \epsilon_0\epsilon_r$ is the electric
permittivity.  The planar surfaces exert a hard-wall potential
$V_{\mathrm{wall}}(z) = w(z) + w(h-z)$, where
\begin{equation}
  w(z) =
  \begin{cases}
    \infty, & z < d/2, \\[6pt]
    0,      & z \ge d/2,
  \end{cases}
\end{equation}
and $z = h/2$ at the mid-plane.

\subsection{Monte Carlo Simulations}
Our simulations are conducted under $\psi$-control, where
$\psi$ is the electrode surface (Donnan) potential.  Choosing $\psi$ $a priori$ 
will cause a corresponding surface charge density in the electrodes. In this
way $\sigma$ vs $\psi$ relations can be obtained for both
bare electrodes (reference) and electrodes grafted
with titratable polymer chains.

The simulation model consists of two parallel, infinite flat surfaces which are perfectly conducting
separated by a distance $h = 250\,\text{\AA}$.  These are immersed
in the electrolyte described above.  We employed periodic boundary
conditions utilising Ewald summations.  The latter used a 3D minimum-image
cutoff in real space, with a unit cell for which $L_x = L_y = 250\,\text{\AA}$
and $L_z = 2h = 500\,\text{\AA}$ ($L_\alpha$ denotes
the extent of the simulation box along the $\alpha$ axis).  Note that
this unit cell also includes image charges, which arise due to the
perfectly conducting walls.  The number of reciprocal wave
vectors along the confinement direction was twice that in the lateral
dimensions ($k_z^{\max} = 2k_x^{\max} = 2k_y^{\max} = 10$).  A total
of $1385$ reciprocal vectors were included in the Fourier component,
and the Ewald splitting parameter was set to $\alpha = 2.52\times
10^{-2}$ (corresponding to $6.3/L_x$) to ensure rapid convergence of
the sums.  Simulations were performed using the Grand Canonical
Monte Carlo (GCMC) method, adopting Metropolis sampling including 
single-ion insertions with effective single-ion electrochemical potentials that depend on the
$a priori$ chosen Donnan potential.  More details can be found in
Ref.~\cite{Janne2020}.

Titration of grafted monomers was treated within the constant-pH Monte
Carlo framework.  At each step, a randomly selected titratable monomer
was proposed to switch between protonated and deprotonated states, with
a concomitant change in its charge valency $\Delta Z$.  The change in
total energy due to a protonation switch, $\Delta U$, was computed as
\begin{equation}
  \Delta U = \tfrac{1}{2}\Delta U_C + \Delta U_{\mathrm{pH}},
  \label{eq:deltaU}
\end{equation}
where $\Delta U_C$ is the change in the full 3D Ewald energy,
including self-energy and Donnan potential contributions associated
with the change of charge, and the titration energy
$\Delta U_{\mathrm{pH}}$ is given by
\begin{equation}
  \Delta U_{\mathrm{pH}} = \ln(10)\,(\mathrm{pH} - \pKa)\,e\Delta Z.
\end{equation}
The factor $1/2$ in Eq.~(\ref{eq:deltaU}) results from a correction to
the image-charge interaction included in the full Ewald energy, this is
described in detail in Ref.~\cite{Janne2020}.
The titration term accounts for the coupling between the local
ionization state and the external pH reservoir.  Accepted titration
moves update both the local charge on the monomer and the global
electrostatic energy of the system.  These titration moves were
performed in combination with configurational updates of the grafted
chains using local crankshaft and end-rotation Monte Carlo moves.  In
a crankshaft move, an internal monomer of the chain was randomly
chosen and the corresponding chain segment was rotated about the local
axis defined by its adjacent bond vectors.  This procedure preserves
bond lengths while modifying the local chain orientation, thereby
efficiently sampling intrachain conformations.  Dissociated ions
underwent grand canonical moves as well as standard displacement moves.

\subsection{Classical Density Functional Theory}
Similar to the simulations, the classical Density Functional Theory (cDFT)
we use in this study uses $\psi$-control.  Here we assume the 
system is uniform in the the planes parallel to electrode
surfaces, and all particle densities in the theory only 
vary in the $z$-coordinate, perpendicular to 
the electrode plane.  
 The end monomer of each chain is
grafted to the electrode wall via a grafting potential
$\phi_g$, independent of its instantaneous charge state.

In the slit geometry,
the grand potential per unit area
$\omega$ for grafted $R$-mer chains with titratable monomers
is
\begin{equation}
  \begin{split}
    \beta\omega\bigl(T,\phi_g,\mu_c,\mu_a,\mathrm{pH},\psi;
    [N(\mathbf{Z}),\alpha(z),n_c(z),n_a(z)]\bigr) = \quad\\
    \int N(\mathbf{Z})\bigl(\ln[N(\mathbf{Z})]-1\bigr)\,d\mathbf{Z}
    + \beta\int N(\mathbf{Z})\,V_b(\mathbf{Z})\,d\mathbf{Z} \\
    +\int \Bigl[n_c(z)\bigl(\ln[n_c(z)]-1\bigr)
               +n_a(z)\bigl(\ln[n_a(z)]-1\bigr)\Bigr]\,dz \\
    +\beta\int{\cal F}_{HS}^{(ex)}(n_m,n_c,n_a)\,dz
    +\frac{\beta e}{2}\int\Psi(z)\,q(z)\,dz
    -\beta\int\bigl(\mu_c n_c(z)+\mu_a n_a(z)\bigr)\,dz \\
    +\beta\int n_m^e(z)\,\phi_g\,dz + \beta\int V_{wall}(z)\bigl(n_m(z)+n_c(z)+n_a(z)\bigr)\,dz +\\
    +\int n_m(z)\Bigl[\alpha(z)\ln\alpha(z)
                      +(1-\alpha(z))\ln(1-\alpha(z))
                      +\ln(10)\,(\mathrm{p}K_a-\mathrm{pH})\,\alpha(z) \Bigr]\,dz \\
    + \int (n_m(z)\alpha(z)^2+n_c(z)+n_a(z)) V_{image}(z,h) dz    
  \end{split}
  \label{eq:omega}  
\end{equation}
where $\mathbf{Z}=z_1\ldots z_R$,
$N(\mathbf{Z})$ is the polymer density distribution, and
$\Psi(z)$ is the local mean-field potential at $z$,
including the Donnan potential $\psi$, but excluding contributions
from image charges, which are handled separately (see below).
$V_b$ is the bond potential which ensures that connected monomers
are separated a bond distance $b=4${\AA}.
The anion and cation densities are denoted $n_a(z)$ and $n_c(z)$,
with chemical potentials $\mu_a$ and $\mu_c$ (equal in the
restricted primitive model, RPM, adopted here).
The total monomer density due to the grafted chains is
denoted $n_m(z)$, and $\alpha(z) \in [0,1]$ is the local
degree of deprotonation (ionisation).  The density of charged
(deprotonated) monomers is accordingly $\alpha(z)\,n_m(z)$,
while the density of neutral (protonated) monomers is
$(1-\alpha(z))\,n_m(z)$. 
The net charge
distribution is
\begin{equation}
  q(z) = e\bigl(n_c(z) - \alpha(z)\,n_m(z) - n_a(z)\bigr)
         + \sigma\bigl(\delta(z)+\delta(h-z)\bigr).
\end{equation}
where $\sigma$ is the (uniform) surface charge density.
Excluded volume effects are accounted for by
${\cal F}_{HS}^{(ex)}(n_m,n_c,n_a)$, using the Generalised
Flory-Dimer approximation \cite{Forsman:2004b}.

The image charge interactions are handled
by the external potential $V_{image}(z,h)$, which (approximately) measures the
interaction of a monovalent charge with its multiple
self-images, which in turn are partly screened by images from
ions of the opposite charge. 
In our previous work, \cite{Szparaga:15} we suggested a simple 
approach to estimate the net interaction.
With the conducting surface being located at $z = 0$, a positive charge
at $z$ will be attracted by its negative self-image, at $-z$, 
with the same $(x,y)$ coordinates as the real charge.
We assume that this image charge is neutralized by a
uniformly charged circular disc of radius $R_d$, 
centred at the position of the image charge and oriented
parallel with the surface. 
This can be viewed as a projection of the cylindrically symmetric ion atmosphere
onto the plane containing the self-image charge, generating a charged circular disc, at $-z$.
For a given choice of $R_d$, we can then write the approximate ``image potential'', $V_{image}(z,h)$,
for a monovalent charge as:
\begin{eqnarray}
\beta V_{image}(z,h) &  \approx  & \beta V_{self}(z,h) \nonumber \\
&  & +\frac{l_B}{R_{d}^2}
\sum_{k=1,3..}^{k_{max}}[\left(R_d^2+((k+1)h-2z)^2\right)^{1/2} \nonumber \\
&  &+\left(R_d^2+((k-1)h+2z)^2\right)^{1/2}-2kh
]
\nonumber \\
&  &  - \frac{2l_B}{R_{d}^2} \left( \sum_{m=2,4...}^{m_{max}} \left(R_d^2+(mh)^2\right)^{1/2}-mh \right)
\label{eq:Vimlimit}
\end{eqnarray}
where $V_{self}(z,h)$ is the unscreened self-image interaction:
\begin{equation}
\beta V_{self}(z,h)  \approx -\frac{l_B}{2} \sum_{l=1}^{l_{max}} [\frac{1}{2(h(l-1)+z)}+\frac{1}{2(hl-z)}-\frac{1}{lh}]
\end{equation}
where $k_{max}$, $m_{max}$ and $l_{max}$ are upper limits that in principle should be infinite but
in practise are chosen as ``very large'' (the sums only need to be evaluated once per cDFT calculation). 

A suitable choice for the long-range limit of $R_d$ can be found by utilizing
established general relationships. If we by  
$p_{vdW}(0)$ denote the zero frequency van der Waals (vdW) pressure between
two flat conducting surfaces interacting across a vacuum slab, the long-range limiting decay is
$\beta p_{vdW}(0) = -\zeta(3)/(8\pi h^3)$, where $\zeta(n)$ is the Riemann zeta function\cite{Mahanty:1976,Kjellander:1987}.
In the presence of an ionic fluid, this interaction should be screened, 
resulting in a non-algebraic decay \cite{Mahanty:1976}. Now, dielectric fluctuations are ignored in the
Primitive Model, but we can capture the screening via image charge interactions, using an appropriate
value of $R_d$\cite{Szparaga:15,Lu:18,Lu:21}. As we demonstrate in the Appendix, the value of $R_d$ that
cancels $p_{vdW}(0)$ at long range is:
\begin{equation}
  R_d = \sqrt{8}/\kappa
\end{equation}
where $1/\kappa$ is the Debye screening length, which for our monovalent salt can be written as:
\begin{equation}
  \kappa^2 = 8\pi\l_B n_s 
\label{eq:kappa}
\end{equation}
where $n_s$ is the bulk salt concentration. This long-range cancellation has been
numerically verified for a wide range of systems\cite{Szparaga:15,Lu:18,Lu:21}.

The last line of Eq.~(\ref{eq:omega}) is the titration free
energy of the grafted layer.  The first two terms,
$\alpha(z)\ln\alpha(z)+(1-\alpha(z))\ln(1-\alpha(z))$, are the mixing
entropy of protonated and deprotonated states, while the third
term, $\ln(10)\,(\mathrm{p}K_a-\mathrm{pH})\,\alpha(z)$, couples
the local degree of ionisation to the external pH reservoir via
the Henderson-Hasselbalch relation.  Minimisation of
$\omega$ with respect to $\alpha(z)$ at fixed monomer density
$n_m(z)$ yields the local Donnan-shifted Henderson-Hasselbalch
equation
\begin{equation}
  \frac{\alpha(z)}{1-\alpha(z)}
  = 10^{\,\mathrm{pH}-\mathrm{p}K_a}\, e^{\bigl(\beta [e\,\Psi(z)-2\alpha(z) V_{image}(z,h)]\bigr)},
  \label{eq:HH}  
\end{equation}
which shows that the charge carried by the grafted polymer depends
self-consistently on both the solution pH and the local
electrostatic potential.
A positive local potential promotes deprotonation while a negative potential
suppresses it. This charge-regulation mechanism is the
key feature that distinguishes the titratable polymer grafted electrode from
permanently charged and fully neutral grafts. 

The equilibrium grand potential per unit area, $\omega_\mathrm{eq}(h)$, was obtained
by coupled Picard iterations over the polymer density
distribution $N(\mathbf{Z})$, the local ionisation field
$\alpha(z)$, and the simple salt ion profiles $n_c(z)$, $n_a(z)$,
using well-established polymer-cDFT methods
\cite{Woodward:1991,Xie:2016}.

\section{Results}

Our results address two distinct energy harvesting processes.  The
first examines blue energy arising from the salinity and pH gradient between
river water and seawater, while the second focuses on energy recovery
from pH neutralization of industrial wastewater.

\subsection{Energy harvested from a salinity difference}
In this section we explore how grafting titrating polyelectrolyte onto the
electrode affects energy recovery using CapMix.  We will examine the sensitivity of
the results to the: polyelectrolyte $\pKa$; grafting density; river pH, and polymer
chain length.  The harvested energy is compared for bare and grafted
electrodes at salt concentrations of $24\,\mathrm{mM}$ (river) and
$600\,\mathrm{mM}$ (sea).  Since seawater has an average pH of
approximately $8.2$, this value is used throughout our calculations.  River
acidity varies in practice, thus we consider a number of values for this quantity.  We shall 
adopt $\mathrm{pH} = 6.5$ as a reference, which is typical of many
rivers.  The grafted chains have a degree of polymerization
$R = 6$ and a grafting density of $\rho_g = 0.032\,\text{\AA}^{-2}$,
uniformly distributed on the electrode surface.  

\begin{figure}[H]
\centering
\includegraphics[scale=0.5]{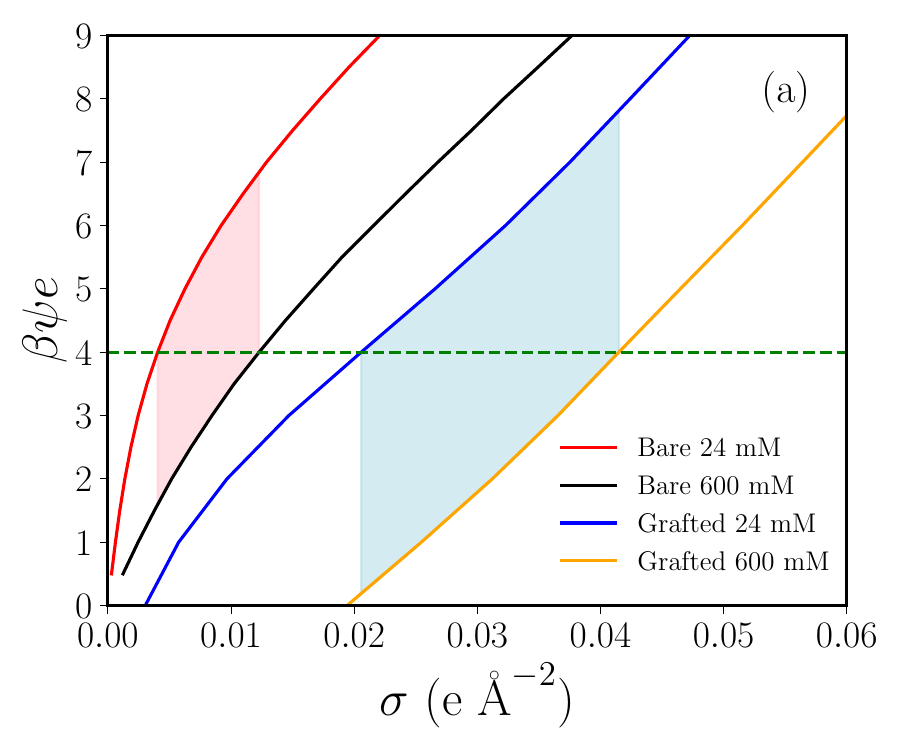}
\includegraphics[scale=0.5]{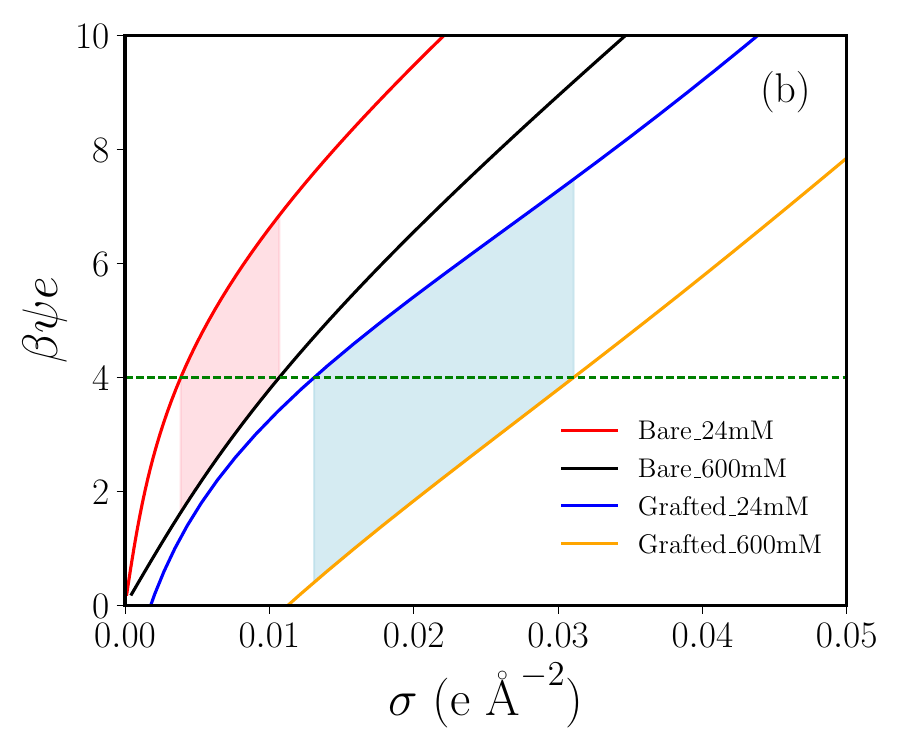}
\caption{Surface (Donnan) potential vs.\ charge curves for bare and
  grafted electrodes at $24\,\mathrm{mM}$ and
  $600\,\mathrm{mM}$ salt concentrations.  The green dashed line
  denotes the external applied potential, $\beta\psi_{\mathrm{ext}}e = 4.0$.
  Shaded regions indicate the corresponding energy harvested per unit
  area for bare (red) and grafted (blue) electrodes.  The river pH is
  assumed to be $6.5$. \newline
  (a) Simulation results. \newline
  (b) cDFT predictions.}
\label{fig1}
\end{figure}

Figure~\ref{fig1} displays the $\psi$ vs $\sigma$ plots of electrodes
with and without grafted titrating polymers with $\pKa = 8.2$.  
The results for fresh- and seawater, $24\,\mathrm{mM}$ and $600\,\mathrm{mM}$,  demonstrate
that for a given surface charge, lower ionic strength yields a higher
potential, owing to the expansion of the electrical double layer (EDL).  The shaded region
(described by the dashed line) between the seawater and freshwater 
curves gives the energy per unit area harvested  for $\beta\psi_{\mathrm{ext}}e = 4.0$
by the thermodynamic cycle described earlier.  This is explicitly given by,
Eq.~(\ref{eq1.1}). 

Where grafted polymers are present, the ionizable monomers titrate in
response to the local electrostatic potential and pH. 
The potential contribution includes surface polarization, as modelled by image
charges, leading to a finite surface charge density even at zero
surface potential.  The coupling between ionization equilibrium, electrostatic
interactions, and confinement leads to a significant modification of
the electrical double layer structure.  This is reflected in an increased
separation between the two branches of the
$\psi$ vs $\sigma$ curves, compared with bare electrodes, 
and a corresponding enhancement of the
energy harvested per thermodynamic cycle. Comparisons with 
cDFT predictions, displayed in Figure \ref{fig1}(b)
indicate that despite its mean-field nature, the cDFT
is reasonably accurate.  The cDFT results support the conclusion that grafting titrating
polyelectrolyte onto the electrode surfaces can significantly enhance the harvested energy. The
higher capacitance in the ocean environment and the greater
polyelectrolyte charge due to the deprotonation is reflected in 
the ion distributions, and the charge density profiles.
\begin{figure}[H]
\centering
\includegraphics[scale=0.5]{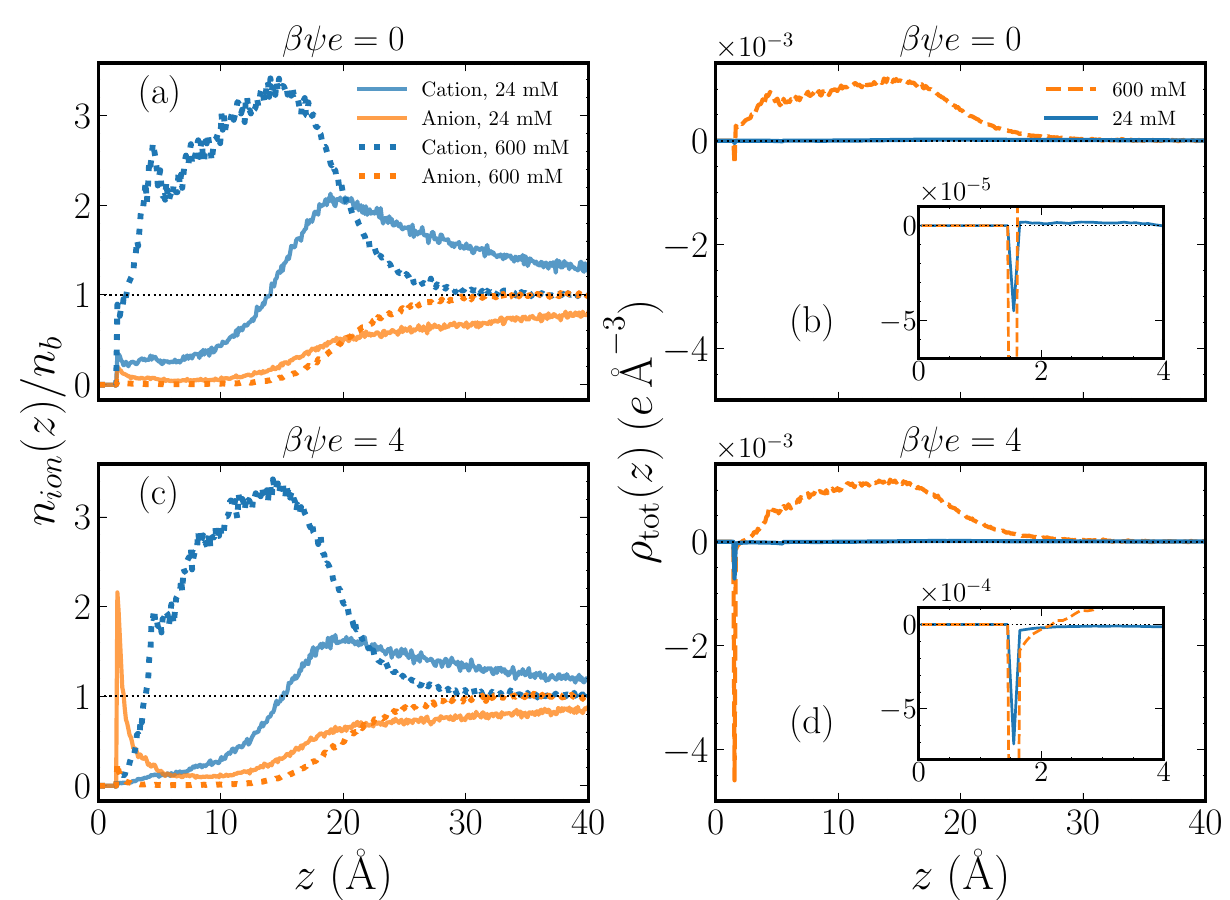}
\caption{Simulated ion distributions and
  charge density profiles. River and ocean conditions are
  identical to those in Figure~\ref{fig1}.
  \\
  (a), (c) Ion density profiles at a reduced Donnan potential of $0$ (a) and $4$ (c). \\
  (b), (d) Total charge density profiles, $\rho_{tot}(z)$,  at a reduced Donnan potential of $0$ (b) and $4$ (d). \\}
\label{fig2}
\end{figure}
Examples are provided in Figure \ref{fig2}, where we note an accumulation of cations near
the surface, even at zero surface potential. This cation response is essentially due to the
negatively charged monomers of the polyelectrolyte grafted to the
electrodes. In seawater, the monomers are almost all charged, due to the
strong self-image attraction which is only weakly screened. In other words, a
negatively charge monomer will effectively form a dipolar dimer with
its (positive) self-image, and any screening cations from the electrolyte are 
repelled due mainly to steric interactions. 
An increase in the surface (Donnan) potential generates an even 
stronger accumulation of cations in seawater, but the effect is 
perhaps smaller than one might anticipate. 

\begin{figure}[H]
\centering
\includegraphics[scale=0.5]{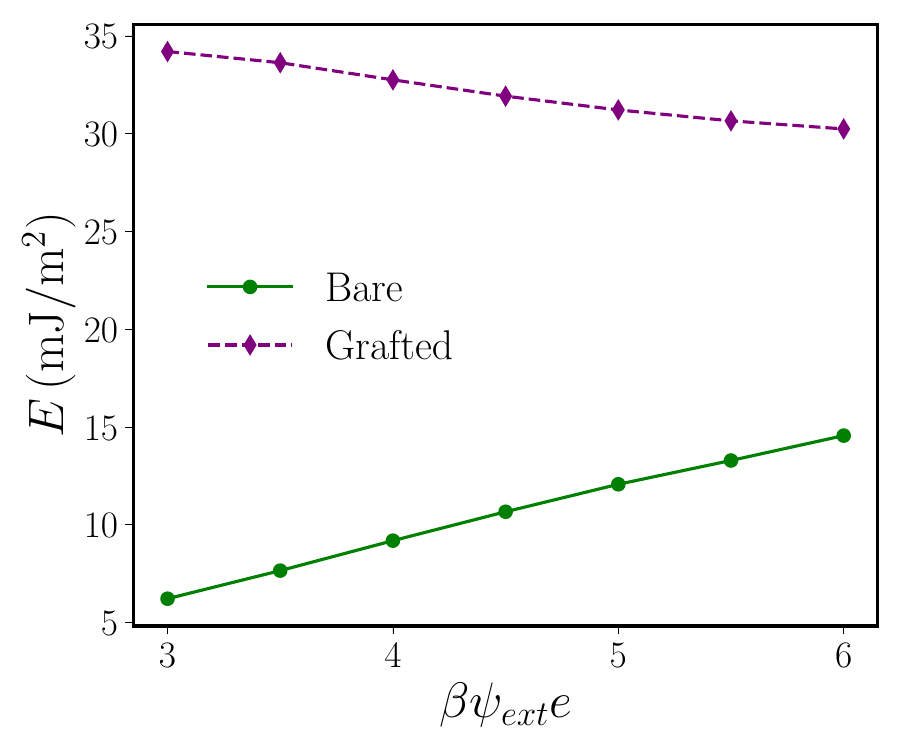}
\caption{Simulated energy extraction per unit surface area per cycle, as a function
  of the applied external potential ($\psi_{\mathrm{ext}}$) for bare
  (green solid line) and grafted (purple dotted line) electrodes during
  salinity gradient mixing between river water ($24\,\mathrm{mM}$) and
  seawater ($600\,\mathrm{mM}$).  River and ocean conditions are
  identical to those in Figure~\ref{fig1}.}
\label{fig3}
\end{figure}

Figure~\ref{fig3} shows the energy harvested per unit surface area at
different values for the external potential, $\psi_{\mathrm{ext}}$, 
for both bare and grafted electrodes.  For bare electrodes, 
the charge-potential curves for high and low salt
solutions both pass through the origin and increase in 
separation as the surface potential increases.
At low potentials, only a small amount of electrode charge is present, resulting in
a low energy harvest.  At higher voltages, the double layer in
high-salinity solutions incorporate significantly more ions than in
low salinity, widening the gap between the two curves and thereby
increasing the extracted energy.  With grafted electrodes, however,
the charge distribution is dominated by the grafted polyelectrolyte, keeping the
potential gap between low and high salinity nearly constant.  As a
result, the harvested energy is significantly enhanced compared to bare
electrodes.  At high external voltages, the relative advantage of
grafted electrodes decreases but remains substantial.
Complementary cDFT results are also presented.
Despite its mean-field nature, cDFT reproduces the trends observed in
Monte Carlo simulations and further confirms that grafting titrating
polymers onto the electrode surface can significantly enhance the
harvested energy.

\subsubsection{Influence of monomer $\pKa$.}
\begin{figure}[H]
\centering
\includegraphics[scale=0.5]{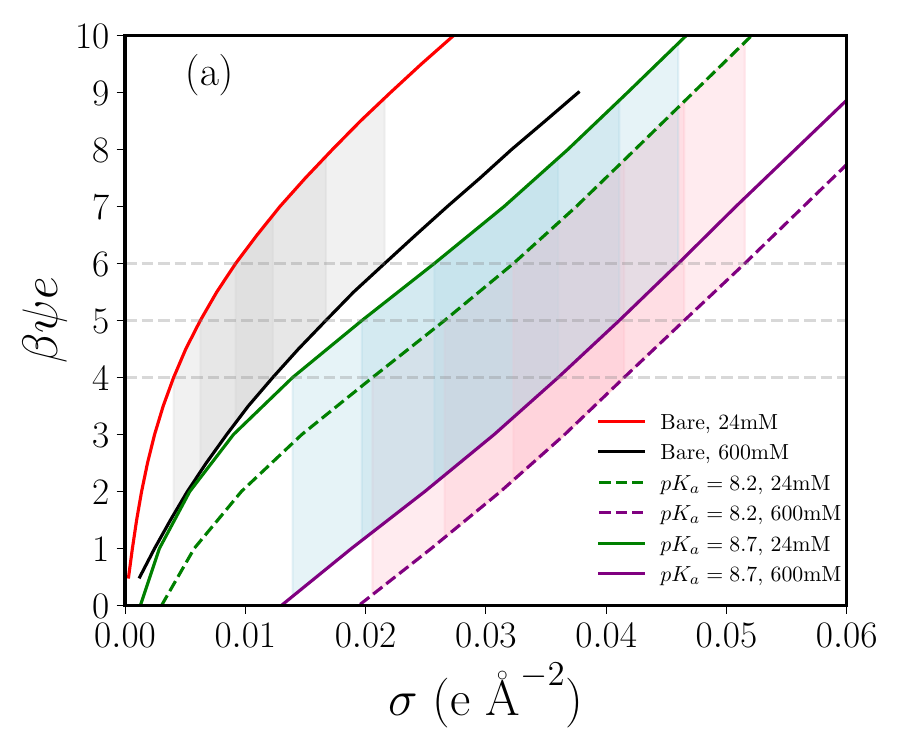}
\includegraphics[scale=0.49]{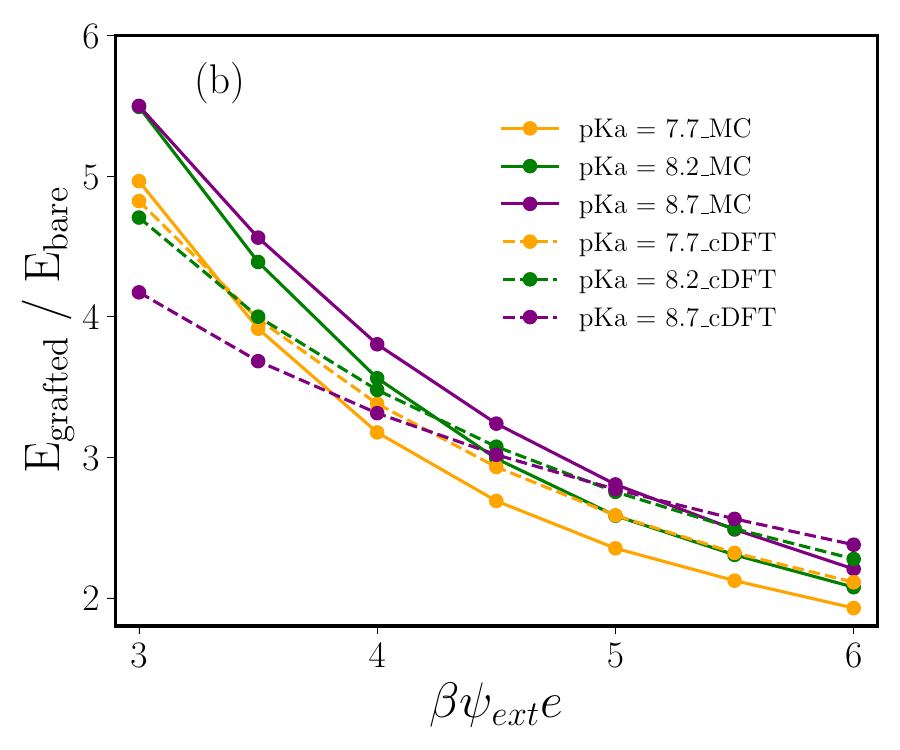}
\caption{(a) Simulated ``blue energy'' loops for bare and grafted electrodes.
  For the grafted case, monomer $\pKa$ values of $8.2$ and $8.7$ are
  compared.  River and ocean conditions are identical to those in
  Figure~\ref{fig1}.  Horizontal dashed grey lines indicate the applied external
  potentials $\beta\psi_{\mathrm{ext}}e = 4.0$, $5.0$, and $6.0$;
  the areas of the shaded regions represent the corresponding
  harvested energy.  \newline
  (b) Ratio of harvested energy obtained with
  grafted ($E_{\mathrm{grafted}}$) and bare ($E_{\mathrm{bare}}$)
  electrodes for the two examined $\pKa$ values. cDFT predictions are
also included (dashed lines).}
\label{fig4pka}
\end{figure}

Figure~\ref{fig4pka}(a) illustrates some effects discussed above
and also the influence of the $\pKa$ of the polyelectrolyte.
The shaded regions show the extracted energy for various electrode
conditions, external potentials, and monomer $\pKa$ values.  In
Figure~\ref{fig4pka}(b), we plot the ratio of the harvested energy with
bare ($E_{\mathrm{bare}}$) and grafted ($E_{\mathrm{grafted}}$)
electrodes.  As already noted, the improvement obtained by grafting
decreases as the external potential increases.  Increasing the monomer
$\pKa$ from $8.2$ to $8.7$ generates a marginal additional improvement
in performance.  At low salinity and weak external potentials, a higher
$\pKa$ means the monomers remain more protonated (less ionized) at the
river pH, amplifying the contrast in ionization state between the two
salinity branches and thereby slightly widening the energy loop.  The
same qualitative response is found at high salinity, but ionic
screening weakens the effect.

The observed increase in harvested energy for $pK_a = 8.7$ 
originates from the stronger $pH$-dependent variation
of the monomer ionization state, during the charging and discharging cycle. In
general, when $pK_a \gg pH$, protonation dominates and the ionizable 
groups remain largely neutral at low and intermediate potentials, whereas
deprotonation and ionization increase as the pH approaches or 
exceeds the $pK_a$. For the $pK_a=8.7$ system, the
grafted polyelectrolytes remain strongly protonated and 
weakly ionized at $pH=6.5$, while at $pH=8.2$ a
substantial increase in deprotonation occurs, leading to 
a significantly more ionized polyelectrolyte brush, 
at comparable potentials.  This difference is further promoted by the stronger 
ionic screening at $pH=8.2$ (seawater).
In contrast, for $pK_a=8.2$, the polyelectrolytes 
are already partially deprotonated at $pH=6.5$, and
therefore the increase in ionization upon changing 
to $pH=8.2$ is comparatively smaller.
Consequently, the $pK_a=8.7$ system undergoes a 
somewhat larger surface charge variation during the blue 
energy loop, which enhances the harvested energy slightly.

\subsubsection{Influence of grafting density.}
\begin{figure}[H]
  \centering
\includegraphics[scale=0.49]{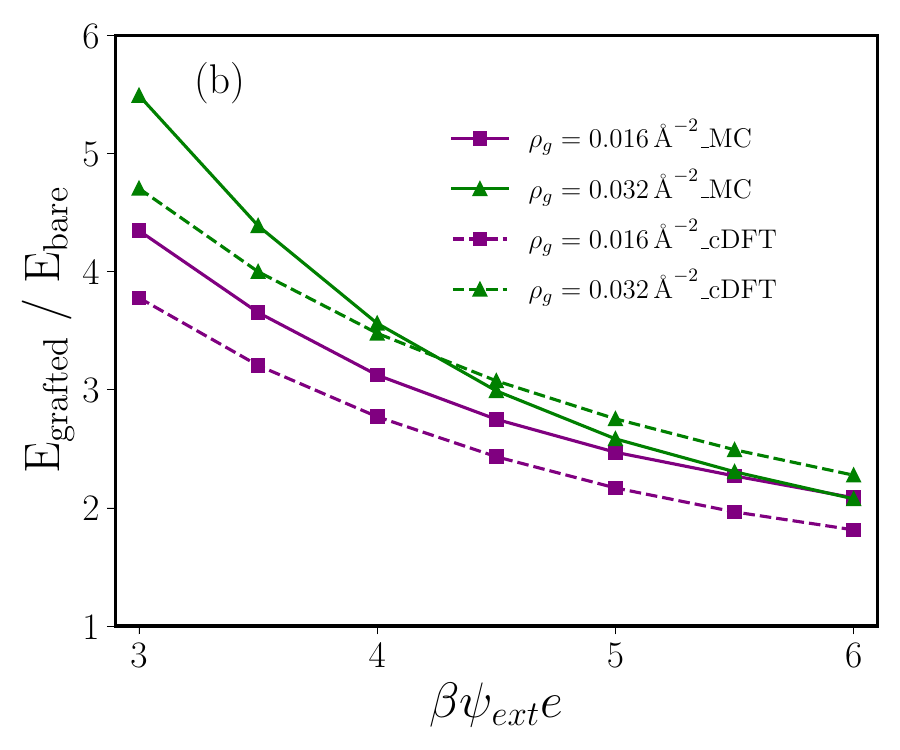}
\caption{Simulated blue energy performance ratios, $E_{\mathrm{grafted}}/E_{\mathrm{bare}}$,
  obtained with our reference (green, $\rho_g = 0.032\,\text{\AA}^{-2}$) and
  reduced (purple, $\rho_g = 0.016\,\text{\AA}^{-2}$) grafting densities,
  for various external potentials. River and ocean conditions are
  identical to those in Figure~\ref{fig1}.}
\label{fig5}
\end{figure}

We have also performed simulations in which the grafting density is reduced
to $\rho_g = 0.016\,\text{\AA}^{-2}$.
As noted in Figure~\ref{fig5}, this leads to 
a slight drop in performance, although the difference is
negligible at high external potentials.  The performance drop at low
external potentials is likely due to the fewer ionizable
groups per unit area. This difference is diminished
at stronger external potentials, where ionization due to titration saturates.

\subsubsection{Impact of river water acidity.}
\begin{figure}[H]
  \centering
\includegraphics[scale=0.49]{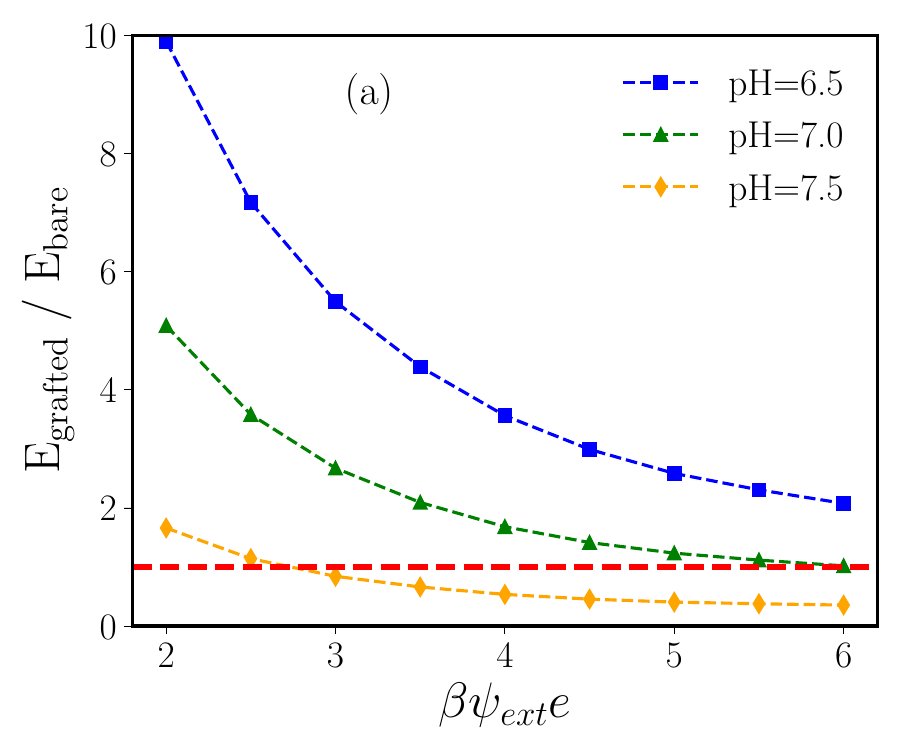}  
\includegraphics[scale=0.49]{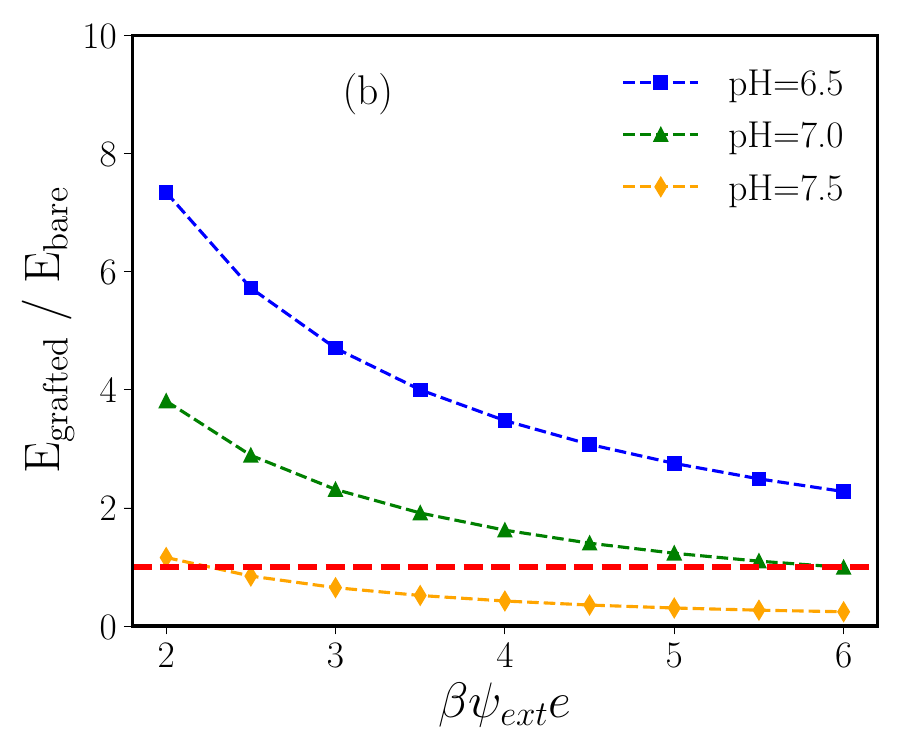}
\caption{Blue energy performance ratios obtained in the presence of
  rivers with different acidity, shown for various external potentials. \newline
  (a) MC simulation results. \\
  (b) cDFT predictions.}
\label{figApHdep}
\end{figure}
In Figure~\ref{figApHdep}(a), we report simulation results
for the harvested energy obtained by
mixing ocean water with freshwater of different acidities---$\mathrm{pH} =
6.5$, $7.0$, and $7.5$---at $\pKa = 8.2$ and various applied
potentials.  As expected, the harvested energy is maximal at the most
acidic condition ($\mathrm{pH} = 6.5$) and decreases with increasing
pH.  At low applied potentials, even the alkaline stream
($\mathrm{pH} = 7.5$) shows modest energy gain due to partial
ionization of grafted chains.  However, with increasing potential, the advantage of grafting
titrating polymers diminishes. This is mainly because the monomers are
strongly ionized even at low salinity.  In principle, this
can be counteracted by using monomers with a higher $\pKa$ value, but
only at the expense of weaker monomer ionization at ocean conditions.
These trends are fully supported by our cDFT calculations shown in Figure~\ref{figApHdep}(a).

\subsubsection{Influence of polymer length.}
\begin{figure}[H]
\centering
\includegraphics[scale=0.5]{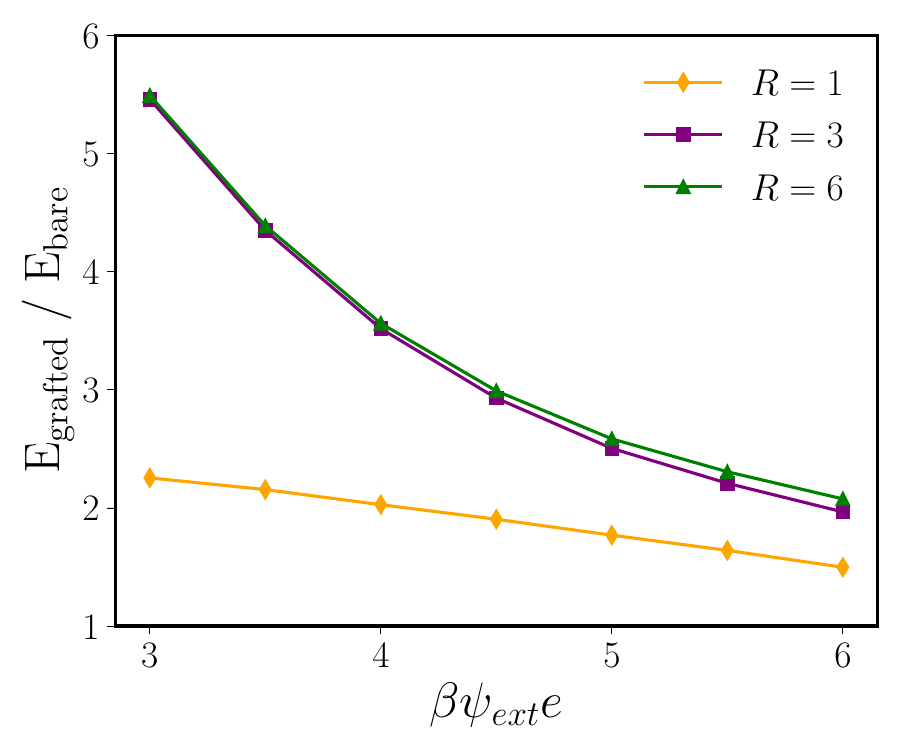}
\caption{Simulated blue energy performance ratios obtained in the presence of
  polymers with various degrees of polymerization $R$.}
\label{chainlength}
\end{figure}
In Figure~\ref{chainlength}, we see that while a purely monomeric layer
provides a weaker improvement than a polymer layer, the chain length
effect saturates rapidly, with negligible changes for $R > 3$.
At our rather high reference grafting density, steric and electrostatic
repulsions force chains into extended conformations, limiting their
flexibility.  Electrostatic interactions are dominated by monomers
near the electrode surface, which strongly couple to their image
charges and induce pronounced ion structuring.  Monomers farther from
the surface provide only a progressively smaller contribution.

\subsection{Energy harvesting from pH neutralization}
The results above show that the charge regulation induced by grafted
titrating polymers can exploit pH differences to enhance energy
extraction.  The same mechanism can be applied to a second,
industrially relevant scenario; harvesting energy from the
neutralization of acidic (or alkaline) wastewater by mixing with
neutral water.  We now apply the CapMix framework to this problem.

\begin{figure}[H]
\centering
\includegraphics[scale=0.5]{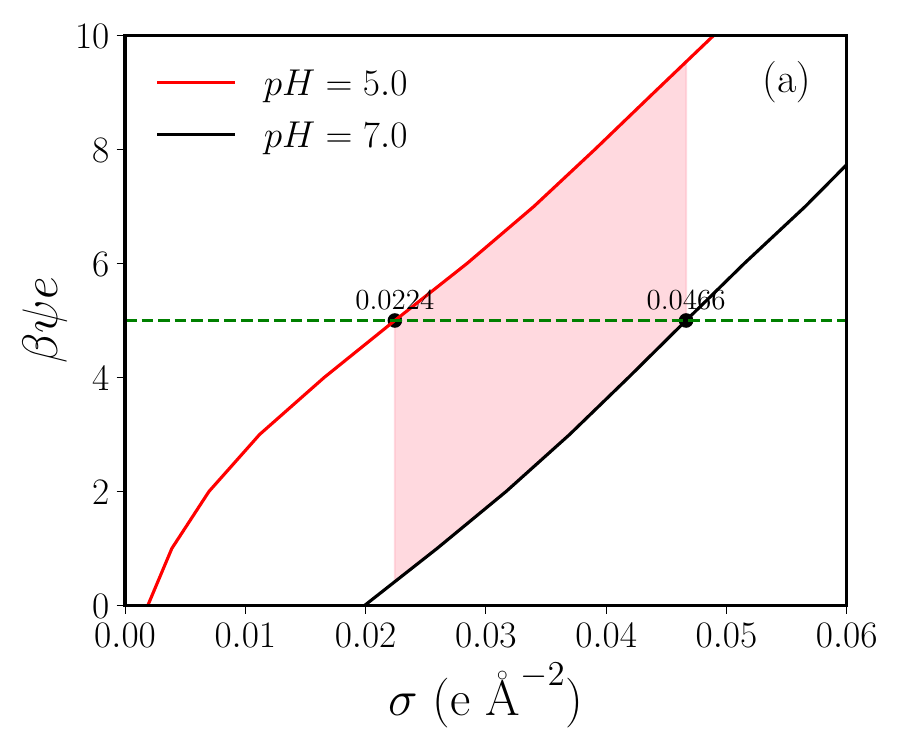}
\includegraphics[scale=0.5]{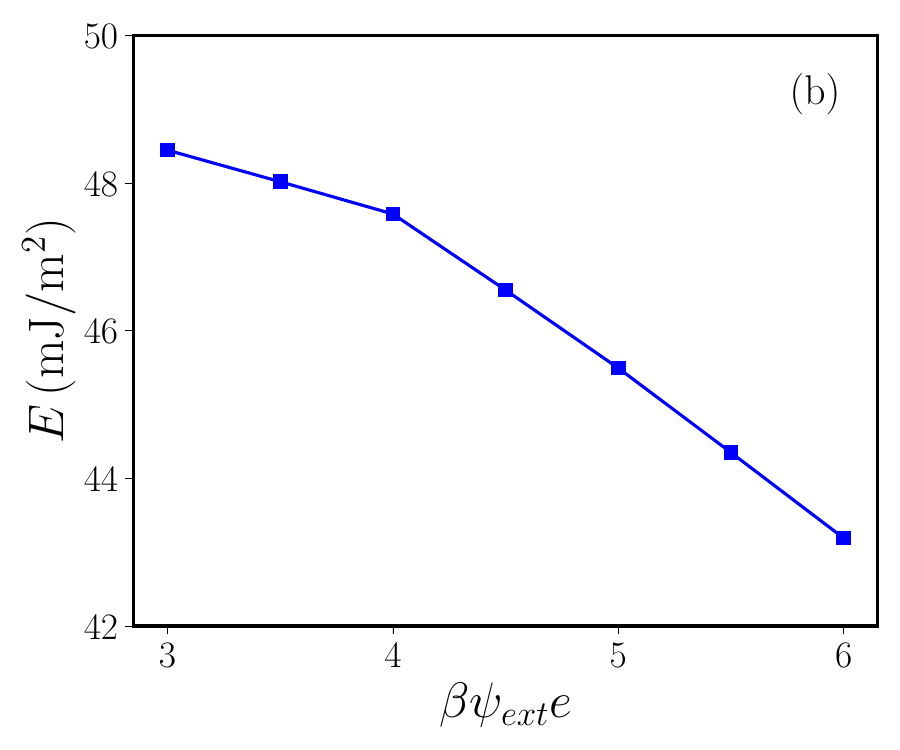}
\caption{CapMix energy harvesting from industrial wastewater. \\
  (a) Simulated blue energy loop at $24\,\mathrm{mM}$ salt for acidic wastewater
  ($\mathrm{pH} = 5$) mixing with neutral water ($\mathrm{pH} = 7$),
  where the titrating monomers of the grafted polymers (6-mers) have a
  $\pKa$ value of $7$.  The dashed line marks the applied potential
  $\beta\psi_e = 5$.  The shaded area represents the energy recovered
  during neutralization per unit area. \newline
  (b) Harvested energy as a
  function of the external potential.}
\label{figphneu}
\end{figure}
Figure~\ref{figphneu}(a) shows the surface potential as a function of
surface charge at a bulk salt concentration of $24\,\mathrm{mM}$,
where acidic wastewater is mixed with neutral water of the same
salinity.  The chosen external potential is given by $\beta\psi_e = 5$.  The shaded area
represents the energy recovered per mixing loop per unit area.  The
harvested energy originates from the charge regulation of the grafted
titrating polymers, which drive ion redistribution upon mixing.  In
Figure~\ref{figphneu}(b), the observed decrease in harvested energy at
higher applied voltages is consistent with the trend in
Figure~\ref{fig3}.  The result demonstrates that a similar
principle can be applied to industrial wastewater treatment, where
both acidic and alkaline waste streams may serve as sources of
extractable energy. While we have conducted these simulations
at a rather arbitrarily chosen salt concentration of
$24\,\mathrm{mM}$, additional simulations
at $600\,\mathrm{mM}$ (not shown) have demonstrated that
the pH dilution harvested energy is effectively
independent of ionic strength (at least with monovalent ions).

\subsubsection{Effect of monomer $\pKa$.}
\begin{figure}[H]
\centering
\includegraphics[scale=0.5]{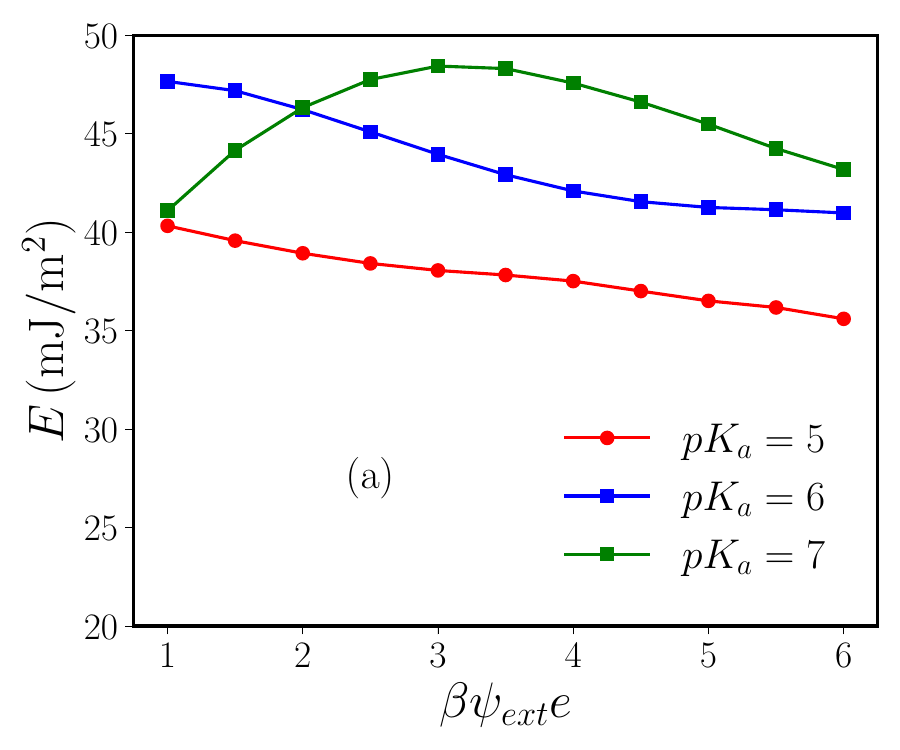}
\includegraphics[scale=0.5]{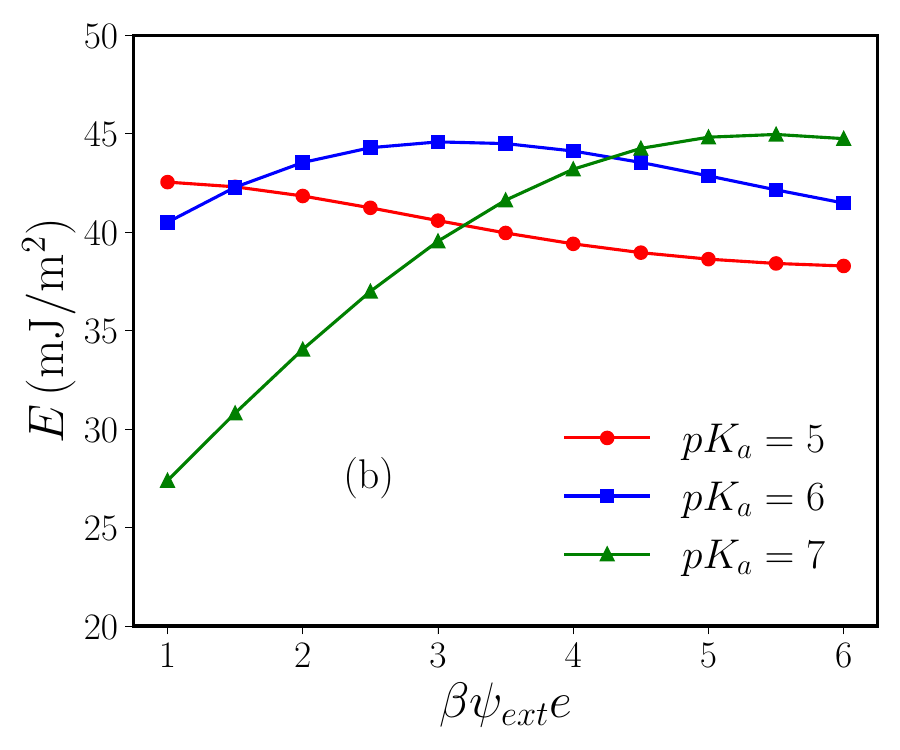}
\caption{Dependence of the pH-mixing energy harvest on $\pKa$, at
  $24\,\mathrm{mM}$ salt concentration.  The extracted energy per unit
  area upon mixing acidic wastewater ($\mathrm{pH} = 5.0$) with
  neutral water is shown for several $\pKa$ values.  The plot illustrates how the titrating
  properties of the surface layer influence the energy recovered during
  the mixing process. \\
  (a) Simulated data.\\
  (b) Data from cDFT calculations
}
\label{figBpka}
\end{figure}

To further examine the role of the polyelectrolyte properties, we analyse the
$\pKa$ dependence in Figure~\ref{figBpka} at a salt concentration of
$24\,\mathrm{mM}$ while mixing water sources at $\mathrm{pH} = 5$ and $7$.
The qualitative trend in harvested energy differs substantially between
low and high $\pKa$.  For $\pKa = 7.0$, the degree of deprotonation
changes strongly between the two sources, resulting in a greater
contrast in graft charge.  The extracted energy is larger at weak
external potentials but decreases with increasing voltage, reflecting
the sensitivity of charge regulation and double-layer saturation near
this $\pKa$.  In contrast, for $\pKa = 8.2$ the change in
deprotonation is much smaller, so the grafts carry only weak charge.
In this case, the applied potential predominantly governs the ion
distribution, and the harvested energy increases monotonically with
voltage.  Comparing the two cases, a low $\pKa$ value provides an
advantage at weak external potentials, although this benefit gradually
diminishes at higher voltages.

\begin{figure}[H]
\centering
\includegraphics[scale=0.5]{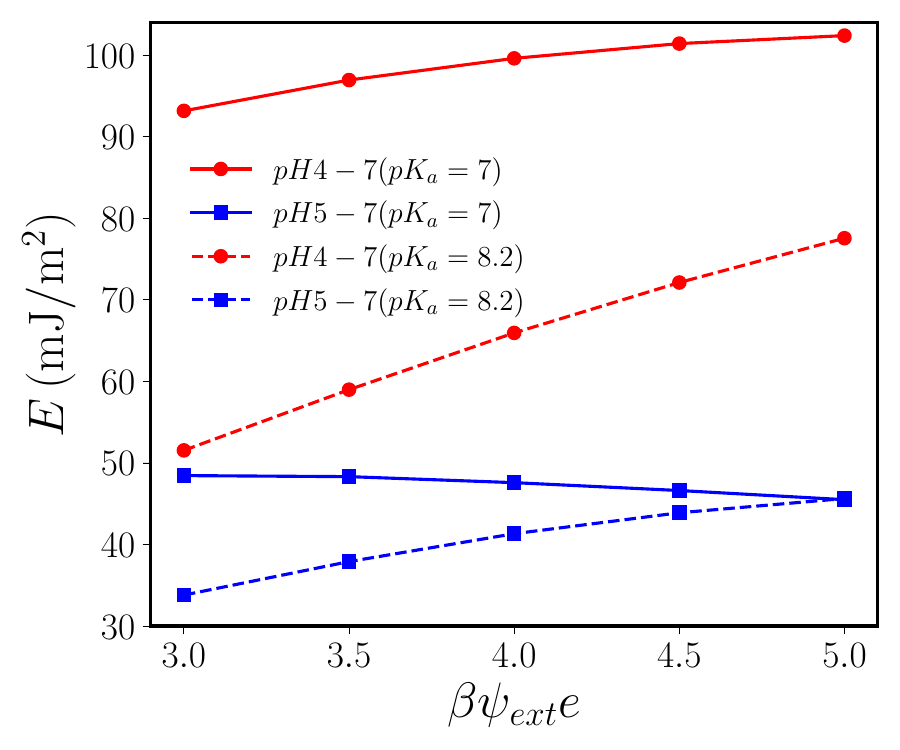}
\caption{Simulated energy extraction per unit area during the neutralization of
  streams from $\mathrm{pH}\,5.0 \to 7.0$ and $\mathrm{pH}\,4.0 \to
  7.0$ for two monomer acidities: $\pKa = 7.0$ (solid curves) and
  $\pKa = 8.2$ (dashed curves). 
  }
\label{figBphpka}
\end{figure}

Figure~\ref{figBphpka} illustrates the overall energy performance for two different wastewater
acidities: $\mathrm{pH} = 4$ and $\mathrm{pH} = 5$.
The dependence of the extracted energy on the external potential is
shown for both acidities, with monomer $\pKa$ values of $7.0$ and
$8.2$.  A more acidic wastewater source offers more harvested
energy, but we also note significant benefits of a higher monomer
$\pKa$ value, especially for the more acidic water source.

\section{Conclusion}
We have demonstrated that grafting titratable polymers onto porous
electrode surfaces yields a substantial enhancement of CapMix energy
recovery.  Using grand canonical Monte Carlo simulations and
complementary cDFT calculations, we have shown that pH-dependent
charge regulation provides a harvested energy at the positive
electrode that is several times higher than that obtained with a
bare electrode.  We have explored the response to variations of
several system parameters---polymer length, grafting density, and
monomer acidity---and found that the energy harvested is rather
insensitive to changes in these parameters within reasonable bounds.
On the other hand, performance is sensitive to the freshwater pH, and the
proposed method is best suited to neutral or acidic rivers.  An
additional application enabled by electrodes covered with titratable
polymers, but not available with bare electrodes, is energy extraction
from the dilution of acidic (or alkaline) water with neutral water.
This strategy offers a promising route to capture electrical energy
from industrial wastewater, transforming streams of acidic or alkaline
effluents into a valuable energy resource while simultaneously
mitigating environmental impact.

\section{Appendix: a general derivation of  $R_d=\sqrt{8}/\kappa$}
Here, we relax the constraint of ions being monovalent. 
Consider a tagged ion of charge \(q_i=Z_i e\) in a bulk electrolyte.
Assuming linear response, the charge density $\rho_i$ of its screening atmosphere is
\begin{equation}
  \rho_i(r)
  =
  -q_i\,\frac{\kappa^2}{4\pi}\frac{\exp(-\kappa r)}{r},
  \label{eq:debye-cloud}
\end{equation}
where \(\kappa\) is the inverse Debye length.  The atmosphere has total
charge \(-q_i\):
\begin{equation}
  \int \rho_i({\bf r})\,d{\bf r}=-q_i .
\end{equation}
Its normalized radial second moment is
\begin{align}
  \langle r^2\rangle_{proj}
  &=
  \frac{1}{-q_i}\int r^2\rho_i({\bf r})\,d{\bf r} \notag\\
  &=
  \kappa^2\int_0^\infty r^3 e^{-\kappa r}\,dr
  =
  \frac{6}{\kappa^2}.
  \label{eq:radial-second-moment}
\end{align}
Since the distribution is isotropic, we have $<x^2>=<y^2>=<z^2>=<r^2>/3$. Hence,
the 2D-projected moment, $<r^2>_{proj}$ becomes 
\begin{equation}
  <x^2+y^2> \equiv <r^2>_{proj} =   \frac{4}{\kappa^2}.
\end{equation}
For a uniformly charged disc, with radius $R_d$, we obtain:
\begin{equation}
  <x^2+y^2>_{disc} = \frac{\int_0^{R_d} \rho_0^2 2\pi \rho_0 d\rho_0}{\int_0^{R_d} 2\pi \rho_0 d\rho_0} = \frac{R_d^2}{2}
\end{equation}
where we have added an index ``0'' to separate the 2D integration variable from the charge density. 
By equating $<r^2>_{proj}$ and $<x^2+y^2>_{disc}$, we obtain
\begin{equation}
  R_d = \frac{\sqrt{8}}{\kappa}
\end{equation}
We now proceed to demonstrate that this choice also leads to the correct
limiting interaction pressure, i.e. one that cancels $p_{vdW}(0)$.
Let $P(\rho_0)$ denote a normalized, circularly symmetric counterion distribution on the image plane, and
define the moment
$M_2 \equiv \int \rho_0^2 2\pi\rho_0 P(\rho_0) d\rho_0$.
Then we note that
for a large separation, we have:
\begin{equation}
  \int \frac{P(\rho_0)2\pi\rho_0}{\sqrt{L^2+\rho_0^2}} d\rho_0 \approx \frac{1}{L}-\frac{M_2}{2L^3}    
  \label{eq:multipole}
\end{equation}
The monopole $1/L$ term cancels, since the overall distribution is electroneutral.

For two conducting walls, separated by $h$, the wide slab limiting expression for
the $m$th order reflection distance $L_m$ becomes:
\begin{equation}
  L_m=2mh, \hspace{25mm} m=1,2, ...
  \label{eq:reflection-distance}
\end{equation}
(the local $z$-dependent offsets can be neglected at large separations). This leads us to the
following expression for the interaction free energy per unit area, $\omega_i(h)$ of a single ion of 
species $i$:
\begin{equation}
  \beta\omega_i \approx Z_i^2l_B\sum_{m=1}^\infty \frac{M_2}{2(2mh)^3} = Z_i^2l_B\frac{M_2\zeta(3)}{16 h^3}
\end{equation}
where we have utilized the relation $\zeta(3) = \sum_{m=1}^\infty m^{-3}$.
In a very wide slit, we can approximate $\int_0^h n_i(z)dz \approx n_i(b)h$ where
$n_i(b)$ is the bulk density of species $i$.
We then arrive at the following expression for the image grand potential per unit area, $\Omega_{im}/A$:
\begin{equation}
  \frac{\beta \Omega_{im}}{A} \approx \sum_i n_i(b) h \beta \omega_i = \frac{l_B M_2 \zeta(3) \sum_i n_i(b) Z_i^2}{16 h^2}
\end{equation}
Using $\kappa^2 = 4\pi l_B \sum_i n_i(b)Z_i^2$ we can simplify this as:
\begin{equation}
  \frac{\beta \Omega_{im}}{A} \approx \frac{l_B M_2 \zeta(3) \kappa^2}{64\pi h^2} = \frac{R_d^2 \zeta(3) \kappa^2}{128\pi h^2}
\end{equation}  
where the last equality results from $M_2 = R_d^2/2$. This leads us to a wide separation
limit for the image pressure $p_{iPB}$:
\begin{equation}
  \beta p_{iPB} = -\frac{\partial\left({\frac{\beta \Omega_{im}}{A}}\right)}{\partial h} \approx \frac{\kappa^2 R_d^2 \zeta(3)}{64\pi h^3}
\end{equation}
and with $R_d^2 = 8/\kappa^2$ the final result is
\begin{equation}
  \beta p_{iPB} = \frac{\zeta(3)}{8\pi h^3}
\end{equation}
ensuring that the zero frequency van der Waals pressure is perfectly cancelled at large separations.

\begin{acknowledgement}
J.F.\ acknowledges financial support from the Swedish Research
Council.  We are also grateful for access to the Lund University
high-performance computing service, LUNARC.
\end{acknowledgement}

\bibliography{poly}

\end{document}